\documentclass[12pt,draftcls,onecolumn,peerreview]{IEEEtran}
\usepackage {graphicx}
\usepackage{subfigure}
\DeclareGraphicsExtensions{.eps,.pdf,.jpeg,.png}
\usepackage[cmex10]{amsmath}
\usepackage{amssymb}
\usepackage{cite}
\usepackage{bm}
\usepackage{extarrows}
\usepackage{lineno}
\usepackage{mathrsfs}
\title{Capacity of Fading Channels without Channel Side Information}

\author{Xuezhi Yang,~\IEEEmembership{Senior Member,~IEEE }\thanks{This manuscript is currently under the review of \emph{IEEE Transactions on Information Theory}. The first version of the paper was submitted to \emph{Nature} on Sept. 21, 2017 and to \emph{IEEE transactions on Information Theory} on Jan. 20, 2018. }}

\begin{document}
\maketitle


\begin{abstract}
There are currently a plurality of capacity theories of fading channels, including the ergodic capacity for fast fading channels and outage capacity for slow fading channels.  However, analyses show that the outage capacity  is a misconception. In this paper we use  the 1st order Gaussian-Markov process  with coherence coefficient $\alpha$ as the unified model for slow and fast fading channels, the capacity of which without channel side information is studied. We demonstrate that the information rate of a fading channel has a structure that the rate of user message is always accompanied  by a rate of channel information. The formula for the channel information rate is  derived and turns out to be a non-increasing function of $\alpha$. We prove that there is an asymptotically monotonic behavior of the user information rate with respect to $\alpha$ when the input is independent, identically distributed and Gaussian in the high signal to noise ratio regime. It is further  conjectured that the monotonic behavior of the user information rate with respect to $\alpha$ is universal. 
\end{abstract}


\section{Introduction}

Channel capacity, defined as the supremum of all achievable data rates with arbitrarily low probability of error on a channel,  is the fundamental concept in Shannon's information theory \cite{ShannonInformationTheory1948}, which has been guiding the development of communications industry since 1948. 

Let's consider the complex-valued additive white Gaussian noise (AWGN) channel
\begin{equation}\label{AWGNch}Y=X+Z,\end{equation}
where $X,Y,Z$ denote the input signal, output signal and  Gaussian noise respectively. If not otherwise specified, all the complex random variables and vectors in this paper are circularly symmetric. The capacity of (\ref{AWGNch}) is the maximum of the mutual information between $X$ and $Y$ over all distributions on $X$ satisfying the power constraint $EX^2\leqslant\sigma_X^2$, i.e.
\begin{equation} C_{AWGN}=\max_{p(x):EX^2\leqslant\sigma_X^2}I(X;Y),\end{equation} 
where $p(x)$ is the probability density function (PDF) of $X$, and $I(X;Y)$ is the mutual information of $X$ and $Y$. Through further calculation, the channel capacity of the AWGN channel is found to be
\begin{equation}
  C_{AWGN}=\log\left(1+\frac{\sigma_X^2}{\sigma_Z^2}\right) \textrm{bits/symbol},
\end{equation}
where $\sigma_Z^2$ denotes the variance of  $Z$, $\log x$ is the base-two logarithm of $x$. 

The operational meaning of channel capacity is manifested by the channel coding theorem. That is, all rates below capacity are achievable. Specifically, for every rate $R<C_{AWGN}$,  there exists a code whose bit error rate can be arbitrarily small.  Conversely, any code with arbitrarily low probability of error must have a rate $R\leqslant C_{AWGN}$ \cite{ShannonInformationTheory1948, Gallager1968Information, Cover1991Elements}. 

Wireless communications have evolved into a huge industry because of the conveniences provided by its mobility. Fading, caused by reflection  and diffraction of electromagnetic waves, is a common phenomenon in wireless communications. The capacity of fading channels has been attracting  researchers' interests for decades because of both its  theoretical and practical significances.  

Fading in wireless communications is usually  classified into slow fading and fast fading. Slow fading can be caused by events such as shadowing, where a large obstruction obscures the main signal path between a transmitter and a receiver. Fast fading occurs when the signal travels from a transmitter via multiple paths to a receiver  where multiple replicas of the signal are combined together, and there exists a relative speed between the transmitter and the receiver \cite{AndreaWC, DavidTse, Tao2016}.  

In this paper we study the discrete-time fading channel model   
\begin{equation} \label{ModelDMCH1}Y_i=G_iX_i+Z_i, \end{equation} 
where $i$ stands for the discrete time or frequency index, the sequence $\{X_i\}$ is the complex-valued transmitted sequence satisfying the average-power constraint $E[|X_i|^2]\leqslant \sigma_X^2$, where $E[\cdot]$ denotes expectation. The independent  and identically distributed (i.i.d.) and Gaussian noise samples are designated by  $\{Z_i\}$,  $E[|Z_i|^2]=\sigma_Z^2$.   The sequence  $\{Y_i\}$ stands for the complex-valued received signal samples. The sequence $\{G_i\}$ is complex-valued channel gain coefficients.  Even though the inter-symbol interference (ISI), which is common in wireless communications,  is not considered here, the model is applicable in practical systems like the already launched 4G \cite{Dahlman20114G} and the ongoing 5G \cite{TS28212} systems, where orthogonal frequency division multiplexing (OFDM) is used and  ISI is eliminated by cyclic prefix. 

If $G_i$ are independent, the model in (\ref{ModelDMCH1}) is an i.i.d. channel \cite{G2002Capacity, Abou2001The}.   Another specific example of  the model in (\ref{ModelDMCH1}) is the block fading channel \cite{Knopp2000On, Chowdhury2016Capacity}, where the  channel gain $G_i$ remains constant within one time block of length $T_0$, then  changes randomly to another level and remains constant for another $T_0$, and so on. The block fading model is useful in analyzing mobile radio systems which employ techniques such as slow frequency hopping.

The case of perfect CSI known to the receiver only for the model in (\ref{ModelDMCH1}) has been treated by many literatures \cite{Goldsmith1997Capacity, Telatar1999Capacity, 720551} and indeed is rather standard.   In such a case, the capacity is known as 

\begin{equation}\label{ErgodicCapacity}
C_{\textrm{Ergodic}}=E\left[\log\left(1+\frac{|G|^2\sigma_X^2}{\sigma_Z^2}\right)\right],
\end{equation}
where the expectation is taken on $G$, denoting anyone of $G_i$.  

$C_{\textrm{Ergodic}}$ is called \emph{ergodic capacity}. It should be noted that, even though  it equals the average of the rates with different channel gains mathematically,  the transmitter can not use  variable rate codes corresponding to different CSI since they are unavailable to the transmitter.  Actually, a standard single rate Gaussian code is enough to achieve $C_{\textrm{Ergodic}}$ in the Shannon meaning. 

When CSI is known to both the transmitter and receiver, there arises another dimension of freedom as power control, so that \emph{water pouring} principle can be applied in the time domain to achieve a higher capacity than $C_{\textrm{Ergodic}}$ \cite{Goldsmith1997Capacity}.

In a slow fading channel,  the codeword length is usually far smaller than the coherence period of the channel, which is thought to be non-ergodic. Researchers have tried to define capacity for non-ergodic channels \cite{335960}. Let's take slow fading channel as an example,  the channel capacity is thought to be a random variable determined by the channel gain. If the transmitter encodes data at a constant rate, there is a non-zero probability that the decoding error probability cannot be made arbitrarily small, in which case the system is said to be in outage. In such a situation, capacity versus outage is used  \cite{Ozarow1994Information, Telatar1999Capacity, AndreaWC}.  The  $\epsilon$-outage capacity is the rate $R_{\epsilon}$ such that
\begin{equation}
Pr(\log\left(1+\frac{|G|^2\sigma_X^2}{\sigma_Z^2}\right)<R_{\epsilon})<\epsilon, 
\end{equation} where $Pr(\cdot)$ denotes probability and $G$ denotes anyone of $G_i$.

In the context of outage capacity, it is interpreted that the capacity of a slow fading channel  in the strict Shannon sense is zero \cite{AndreaWC, DavidTse, 720551}. However, if the outage probability of a slow fading channel with unitary rate is $p$, then it is at least comparable to a binary symmetric channel with crossover probability $p$ and channel capacity $1-H(p)$. Then the capacity of a slow fading channel is definitely NOT zero, which is an indication that the outage capacity theory might be wrong. 

To investigate that, let's revisit the basic concepts of Shannon's theory. In the proof of the channel coding theorem \cite{Gallager1968Information, Cover1991Elements}, a codeword is a stochastic process produced independently in the time domain according to the distribution of the input in the sample space, and so do the noise and channel. The reason we can do that is we assume all these physical stochastic  processes are ergodic.  In the general definition, a random process is ergodic if its time average is the same as its average over the sample space \cite{Walters1982An}. But in information theory, an ergodic process means its time statistics is the same as the statistics over the sample space, which is a much stronger condition required by the channel coding theorem. So, we can conclude that ergodicity is used implicitly as a necessary condition for the channel coding theorem.  In such a circumstance, it is impossible to define a capacity in Shannon's sense for non-ergodic channels. 

For example, the authors of \cite{335960} can not get arbitrarily low error probability in the example in the introduction part, but they fail to realize that there are infinite number of, instead of one,  blocks in one codeword in Shannon's approach. The basic concept of \cite{335960}, information density, is based on the notion that "the distribution of a random variable is random", which is also wrong.   This wrong notion is  embodied in the outage capacity theory, which first deems the channel as an AWGN channel with gain $G_i$,  and then regards  the distribution of  $G_i$  random. However,  the sequence $\{G_i\}$ has a well defined distribution which is NOT random. When using short codewords with a slow fading channel,  the reason you can not get arbitrarily low error probability is not the capacity is zero, but you are not doing it in Shannon's way, in which the codeword should be long enough so that the ergodicity is satisfied. Therefore, with the short codeword approach in a slow fading channel,  we should deem a slow fading channel as many piecewise ergodic AWGN channels from the theoretical perspective if we want to call  the expression \begin{math}
\log(1+|G_i|^2\sigma_X^2/\sigma_Z^2)
\end{math} a capacity. This is already an approximation and should not be pushed further.  The traditional \textit{outage capacity} should be more properly called \textit{outage rate}, because it is not a capacity at all in the Shannon sense. 

The ergodic capacity in (\ref{ErgodicCapacity})  relies on the full knowledge of CSI at the receiver. However, it is impossible for the receiver to obtain perfect CSI since there exists inevitable error  in channel estimation. It is demonstrated in \cite{Lapidoth2002Fading} that the capacity is sensitive to channel estimation  error, not to mention channel estimation is totally meaningless when $G_i$ are independent.   Besides, if we consider the resources for channel estimation and data transmission as a whole, no CSI is a more  reasonable assumption.

A series of fundamental contributions have been made on the capacity of fading channels without CSI.  It is conjectured by Richters \cite{Richters1967COMMUNICATION} and proved in \cite{Abou2001The} that the capacity achieving distribution for an i.i.d. Rayleigh fading channel  is discrete with a finite number of mass points,  one of which located at zero.  This result also holds for i.i.d. Ricean channel \cite{Gursoy2005The}. A supremum to the capacity of an i.i.d. Rayleigh fading channel  is computed using variational methods in \cite{G2002Capacity}. The capacity of block fading MIMO channel is investigated in \cite{746779}, with one conclusion stating that, as the length of the coherence interval increases to infinity,  the capacity without CSI approaches the capacity with perfect CSI. The notion of unitarily invariant codes are proposed in  \cite{ Hochwald2000Unitary, Zheng2002Communication} for block fading MIMO channels without CSI. The analysis and numerical results for block fading channel in \cite{Chowdhury2016Capacity} suggest that pilot-based demodulations are suboptimal and channel estimation is implicitly performed by a capacity-optimal decoder.

All the above studies are focused on either the i.i.d. fading channel or block fading channel. In this paper,  the 1st order Gaussian-Markov process  with coherence coefficient $\alpha$ is used as the unified model for slow and fast fading channels.  The i.i.d. fading channel and block fading channel are just two special cases under $\alpha=0$ and $\alpha=1$, respectively.  We demonstrate that the information rate of a fading channel has a structure that the rate of user message is always accompanied by a rate of channel information. The formula for the channel information part is  derived and turns out to be a non-increasing function of $\alpha$. A much simpler proof of the conclusion in \cite{746779} is provided based on the analytical framework proposed in this paper.  We prove that the user information rate of a fading channel with i.i.d. Gaussian input asymptotically non-decreases with $\alpha$ in the high SNR regime. It is further conjectured that the monotonic behavior of the  user information rate with respect to $\alpha$ is universal.   

The rest of the paper is arranged as follows. The model of fading channels and definition of capacity are described in Section II, Several theorems are presented and discussed  in section III. Conclusions are given in Section IV.  

 \section{Model and definition}

While there are barely practical channels that vary quickly enough to be modeled as  i.i.d. channel, it is also  oversimplified to model the channel coefficient as a constant in a time block,  because no matter how slowly, the channel  is varying anyway in a moving environment, and the effect of such variations can not just be neglected in many cases. For example, multiple pilot symbols are employed in one resource block in the LTE standard to deal with the channel variation caused by user mobility  \cite{Dahlman20114G}. To this end, correlation is introduced to the channel model in (\ref{ModelDMCH1}) to represent the practical systems more accurately in this paper. Specifically, we assume  $\{G_i\}$   is  a stationary and  Gaussian process with $E\left[|G_i|^2\right]=\sigma^2_G$ and $E\left[G_iG^*_j\right]=\alpha^{|i-j|}\sigma^2_G$ for $i\neq j$, $\alpha\in[0,1]$. Therefore this is a typical Rayleigh fading channel with the coherent coefficient between two adjacent samples being $\alpha$.  The following Gaussian-Markov model \cite{841172}
\begin{equation} \label{Gaussian-Markov}G_{i}=\alpha G_{i-1}+\sqrt{1-\alpha^2}W_i,  \end{equation} 
where $\{W_i\}$ is an i.i.d. and Gaussian process with $E\left[|W_i|^2\right]=\sigma^2_G$, is  a channel with the specified properties. 

We divide the time axis into time blocks with length $N$,  then we can rewrite the model (\ref{ModelDMCH1}) for one time block in the matrix form
\begin{equation}
\label{ModelMatrix}
\bm{Y}=\bm{G}\bm{X}+\bm{Z},
\end{equation}
where 
\begin{equation}\notag\bm{X}=\left[  X_1, X_2,  \cdots, X_N \right] ^T, \end{equation}
\begin{equation}\notag\bm{Y}=\left[ Y_1, Y_2 ,  \cdots,Y_N  \right] ^T,  \end{equation}
 \begin{equation}\notag\bm{Z}=\left[ Z_1, Z_2 ,  \cdots ,Z_N\right] ^T, \end{equation}
 \begin{equation}\notag\bm{G}= \textrm{diag} \left[G_1, G_2, \cdots, G_N \right]. \end{equation}  

Here, $N$ is a positive integer,  diag($\bm{V}$) returns  a  diagonal matrix with the elements of  $\bm{V}$ on the main diagonal. 

Suppose the joint PDF of $\bm{X}$ is $p(\bm{x})$, where $\bm{x}=[x_1, x_2,\cdots,x_N]^T$, the capacity of the fading channel in (\ref{Gaussian-Markov}) is defined as \cite{Gallager1968Information}\cite{Goldsmith1996Capacity}
\begin{equation}
\label{capacityblock}
C=\lim_{N\to\infty}\max_{p(\bm{x})}\frac{1}{N}I(\bm{X};\bm{Y}).
\end{equation}
The maximum is taker over all possible distributions of $\bm{X}$, subject to the power constraint $E|X_i|^2\leqslant \sigma_X^2, i=1,2,\cdots,N$. 

The channels of different time blocks are correlated in general, but we have to assume they are independent in order to use $I(\bm{X};\bm{Y})$ as the measure of information rate. To eliminate the effect of channel correlation of two adjacent time blocks, they should be concatenated as one block. Therefore, the influence of channel correlation across blocks is diminishing  when $\alpha\neq 1$ and $N\to \infty$. 

When  $\alpha=1$, the model in (\ref{Gaussian-Markov}) is non-ergodic. As we discussed before, it is impossible to define a capacity for non-ergodic channel. Actually, a non-ergodic channel means, the claimed distribution is not happening in a realization and thus not applicable. A rational way is to take the happing distribution so that the non-ergodic channel is converted to an ergodic channel and  the Shannon theory is applicable. In such a way, any realization of the model  in (\ref{Gaussian-Markov}) is an AWGN channel with gain $G_0$. 

Even $\alpha=1$, the model in (\ref{ModelMatrix}) is still ergodic since the cross-block independence is assumed. The definition in (\ref{capacityblock}) is still meaningful for (\ref{ModelMatrix}).  It is the supremum of the capacity of (\ref{Gaussian-Markov}) with  $\alpha=1-\epsilon,0<\epsilon\leqslant1$, when $\epsilon\to 0$.

\section{Results and Discussions}

\noindent\textbf{Theorem 1}: If $\bm{G}$, $\bm{X}$ and $\bm{Z}$ are mutually independent, then
\begin{equation}
I(\bm{X};\bm{Y})+I(\bm{G};\bm{X},\bm{Y}) = h(\bm{Y})-h(\bm{Z}).
\end{equation}

\noindent\textbf{Proof}: \begin{eqnarray}
I(\bm{X};\bm{Y}) & = & h(\bm{Y})-h(\bm{Y}|\bm{X})\notag\\
 &=&  h(\bm{Y})-h(\bm{Y}|\bm{X},\bm{G})-I(\bm{Y};\bm{G}|\bm{X})\notag\\
 & = & h(\bm{Y})-h(\bm{GX}+\bm{Z}|\bm{X},\bm{G})-h(\bm{G}|\bm{X})+h(\bm{G}|\bm{X},\bm{Y})\notag\\
 & = & h(\bm{Y})-h(\bm{Z})-h(\bm{G})+h(\bm{G}|\bm{X},\bm{Y})\notag\\ &=&  h(\bm{Y})-h(\bm{Z})-I(\bm{G};\bm{X},\bm{Y}). \square \notag
\end{eqnarray}

The independence assumption means the transmitter knows the statistical property of $\bm{G}$ and $\bm{Z}$ but not their instant realization. In such a case the transmitter uses a predefined code book and power to transmit data without water-pouring based power control. 

If the receiver can decode the user message correctly, it has the knowledge of the channel to the extent of $I(\bm{G};\bm{X},\bm{Y}) $ at the same time. The user message rate $I(\bm{X};\bm{Y})$ is always accompanied by a rate of channel information  $I(\bm{G};\bm{X},\bm{Y})$. The sum of $I(\bm{X};\bm{Y})$ and $I(\bm{G};\bm{X},\bm{Y}) $, which equals $h(\bm{Y})-h(\bm{Z})$, can be defined as the total information rate obtained at the receiver. In an AWGN channel, $I(\bm{X};\bm{Y}) = h(\bm{Y})-h(\bm{Z})$, the total information rate is solely the user message information rate. Then Theorem 1 is about the structure of the information rate of a fading channel. The channel matrix $\bm{G}$ is not necessarily to be diagonal, it can be of any form, so Theorem 1 is valid to any linear channel, including multi-path and multi-antenna channels. 

 It is already observed by numerical analysis in  \cite{Chowdhury2016Capacity} that the CSI can  be obtained at the decoder even without any pilot symbol, which is interpreted by the authors that the channel estimation is implicitly performed by the capacity optimal decoder. Actually in one hand, the receiver can use the user message as the training sequence to get an estimation of the channel explicitly, in the other hand,  the receiver indeed has the channel information to the extent of $I(\bm{G};\bm{X},\bm{Y}) $ once it has $\bm{X}$ and $\bm{Y}$, whether or not it performs the channel estimation from them.   

\bigskip
\noindent\textbf{Theorem 2}:
\begin{equation}
I(\bm{G};\bm{X},\bm{Y}) =  \int p(\bm{x})\log   \det(\bm{I}_N+\frac{\sigma_G^2}{\sigma_Z^2}\bm{A}_N) d\bm{x},
\end{equation}
where $\bm{I}_N$ denotes identity matrix of size $N$,
\begin{equation}\notag
\bm{A}_N=\left[                 
  \begin{array}{ccccc}   
   |x_1|^2 & \alpha x_1 x_2^* & \alpha ^2 x_1 x_3^* & \cdots &\alpha^{N-1} x_1 x_N^*\\  
   \alpha x_2 x_1^* &  |x_2|^2  &  \alpha x_2 x_3^* & \cdots &\alpha^{N-2} x_2 x_N^*\\  
    \alpha^2 x_3 x_1^* &  \alpha x_3 x_2^* & |x_3|^2   &\ddots &\vdots\\
    \vdots &  \vdots & \ddots  &\ddots &\alpha x_{N-1} x_N^*\\
  \alpha^ {N-1}x_N x_1^* &\alpha^{N-2} x_N x_2^*&\cdots&\alpha  x_Nx_{N-1}^* & |x_N|^2
  \end{array}
\right].                 
\end{equation}

\noindent\textbf{Proof}: See Appendix I.$\square$

\bigskip
Furthermore, $I(\bm{G};\bm{X},\bm{Y})$ has the following properties.

\noindent\textbf{Corollary}: 

(1) If $\alpha=0$ and $X_1,X_2,\cdots,X_N$ are i.i.d. with PDF $p(x)$, then
\begin{equation}
I(\bm{G};\bm{X},\bm{Y}) = N \int p(x)\log ( 1+\frac{\sigma_G^2}{\sigma_Z^2}|x|^2) dx.
\end{equation} 

(2) If $\alpha=1$, then
\begin{equation}\label{Corollary2}
I(\bm{G};\bm{X},\bm{Y}) =   \int p(\bm{x})\log ( 1+\frac{\sigma_G^2}{\sigma_Z^2}|\bm{x}|^2) d\bm{x}.
\end{equation} 

(3) Given $p(\bm{x})$, $I(\bm{G};\bm{X},\bm{Y})$ is a non-increasing function of $\alpha$.

\noindent\textbf{Proof}: See Appendix II.$\square$

\bigskip
Then the capacity of the fading channel is
\begin{equation} \label{capacity2}
C=\lim_{N\to\infty}\max_{p(\bm{x})}\frac{1}{N}\left(h(\bm{Y})-h(\bm{Z})- \int p(\bm{x})\log \det(\bm{I}_N+\frac{\sigma_G^2}{\sigma_Z^2}\bm{A}_N) d\bm{x}\right).
\end{equation}

AWGN channel is a special case of fading channels with $I(\bm{G};\bm{X},\bm{Y}) =0$, so our result is compatible with Shannon's classical theory. However it seems  there doesn't exist a closed-form expression of $p(\bm{x})$ that maximizes $I(\bm{X};\bm{Y})$ in general. Furthermore, it would be interesting to investigate its properties  in some special cases.

\bigskip

\noindent\textbf{Theorem 3}: If $\alpha=1$ and the PDF of $G_i$ is $p(g)$, then \begin{equation}C=\int p(g)\log\left(1+|g|^2\frac{\sigma_X^2}{\sigma_Z^2}\right) dg.\end{equation}

\noindent\textbf{Proof}: Since $\alpha=1$, then \begin{eqnarray}I(\bm{G};\bm{X},\bm{Y}) &=&   \int p(\bm{x})\notag\log ( 1+\frac{\sigma_G^2}{\sigma_Z^2}|\bm{x}|^2) d\bm{x}\notag\\&\leq&  \log ( 1+\frac{\sigma_G^2}{\sigma_Z^2}\int p(\bm{x})|\bm{x}|^2 d\bm{x})\notag\end{eqnarray} by Jensen's inequality. We also have
\begin{eqnarray}
 I(\bm{G};\bm{X},\bm{Y})-I(\bm{G};\bm{Y})&=&I(\bm{G};\bm{X}|\bm{Y})\geqslant 0,\notag
 \end{eqnarray} 
thus, \begin{eqnarray}\frac{1}{N} I(\bm{G};\bm{Y})&\leq& \frac{1}{N} I(\bm{G};\bm{X},\bm{Y})\notag\\ &\leq& \frac{1}{N} \log ( 1+\frac{N\sigma_X^2 \sigma_G^2}{\sigma_Z^2}) \to 0\notag\end{eqnarray} when $N\to \infty$. We rewrite $C$ as
\begin{eqnarray}C&=&\lim_{N\to \infty} \max_{p(\bm{x})}\frac{1}{N} \left( h(\bm{Y}|\bm{G})-h(\bm{Z})+I(\bm{G};\bm{Y})-I(\bm{G};\bm{X},\bm{Y})\right),\notag
\end{eqnarray}
then
\begin{equation}\notag
C \leqslant \lim_{N\to \infty} \max_{p(\bm{x})}\frac{1}{N} \left( h(\bm{Y}|\bm{G})-h(\bm{Z})+\log ( 1+\frac{N\sigma_X^2 \sigma_G^2}{\sigma_Z^2})\right),
\end{equation}
and
\begin{equation}\notag
C \geqslant \lim_{N\to \infty} \max_{p(\bm{x})}\frac{1}{N} \left( h(\bm{Y}|\bm{G})-h(\bm{Z})-\log ( 1+\frac{N\sigma_X^2 \sigma_G^2}{\sigma_Z^2})\right).
\end{equation}
Therefore
\begin{eqnarray}
C&=&\lim_{N\to \infty} \max_{p(\bm{x})}\frac{1}{N} \left( h(\bm{Y}|\bm{G})-h(\bm{Z})\right)\\
&=&\int p(g)\log\left(1+|g|^2\frac{\sigma_X^2}{\sigma_Z^2}\right) dg.\notag
\end{eqnarray}

The capacity is achieved when $X_1, X_2,\cdots, X_N$ are i.i.d. and  Gaussian.  $\square$

The result of Theorem 3 is first presented in \cite{746779}. The proof here is much simpler than the upper and lower bound approach in  \cite{746779}.

\bigskip
\noindent\textbf{Theorem 4}: If $\alpha=0$ and $X_1,X_2,\cdots,X_N$ are i.i.d. and  Gaussian, then \begin{math}I(\bm{X};\bm{Y})=I(\bm{G};\bm{Y}).\end{math}

\noindent\textbf{Proof}: \begin{eqnarray}
I(\bm{X};\bm{Y})&=&h(\bm{Y}|\bm{G})-h(\bm{Z})+I(\bm{G};\bm{Y})-I(\bm{G};\bm{X},\bm{Y}).
\end{eqnarray}
Since $\alpha=0$, we have
\begin{equation}
   I(\bm{G};\bm{X},\bm{Y})=N\int p(x)\log ( 1+\frac{\sigma_G^2}{\sigma_Z^2}|x|^2) dx
\end{equation}
by corollary (1). Since $X_1,X_2,\cdots,X_N$ are i.i.d. and  Gaussian, we have
\begin{equation}
 h(\bm{Y}|\bm{G})-h(\bm{Z})=N\int p(g)\log(1+|g|^2\frac{\sigma_X^2}{\sigma_Z^2}) dg.
\end{equation}
It is easy to verify that $I(\bm{G};\bm{X},\bm{Y})= h(\bm{Y}|\bm{G})-h(\bm{Z})$, then we have the conclusion. $\Box$

Theorem 4 shows us an interesting symmetry of $\bm{X}$ and $\bm{G}$ while they are both i.i.d. and  Gaussian.

For a given $p(\bm{x})$, the monotonicity in the channel information $I(\bm{G};\bm{X}, \bm{Y})$ with $\alpha$ gives us an intuition that there might be a monotonicity in  the information rate $I(\bm{X};\bm{Y})$, because $I(\bm{X};\bm{Y})+I(\bm{G};\bm{X},\bm{Y}) = h(\bm{Y})-h(\bm{Z})$, as Theorem 1 indicates. However, since $h(\bm{Y})$ is a function of $\alpha$, the conclusion is not immediate, and a rigorous proof seems quite difficult. Instead, we have the following theorem of asymptotic monotonicity. 

\noindent\textbf{Theorem 5}:  

(1) If $X_1,X_2,\cdots,X_N$ are i.i.d., then  $R_l\leqslant I(\bm{X};\bm{Y})/N\leqslant R_l+\Delta$, where $R_l$ is a non-decreasing function of $\alpha$,  $\Delta=C_{0}-R_{s}$, 
\begin{equation}
\label{ }
C_{0}=\log\left(1+\frac{\sigma_G^2\sigma_X^2}{\sigma_Z^2}\right)
\end{equation}
is the channel capacity of the AWGN channel with the channel gain $\sigma_G$, and
\begin{equation}
\label{ }
R_{s}=\frac{h(\bm{Y}|\bm{G})-h(\bm{Z})}{N}
\end{equation}
is the information rate of user message with the receiver having the perfect CSI.

(2) If $X_1,X_2,\cdots,X_N$ are  i.i.d. and  Gaussian, then
\begin{equation}\Delta= \log\left(1+\frac{\sigma_G^2\sigma_X^2}{\sigma_Z^2}\right)-\int p(g)\log\left(1+|g|^2\frac{\sigma_X^2}{\sigma_Z^2}\right) dg, \end{equation} and \begin{equation}\lim_{\rho\to\infty}\Delta=\gamma\log e, \end{equation} where  $\rho=\sigma_G^2\sigma_X^2/\sigma_Z^2$,  $\gamma$  is the Euler-Mascheroni constant, and $e$ is the Euler's number. 

\noindent\textbf{Proof}: See Appendix III.

Theorem 5 demonstrates that $I(\bm{X};\bm{Y})/N$ lies between two parallel non-decreasing functions, the distance between which is $\Delta$.  Typically,  $\Delta$ is a  small number compared to $I(\bm{G};\bm{X},\bm{Y})/N$. For example, in the case $X_1,X_2,\cdots,X_N$ are i.i.d. and  Gaussian, which is addressed in the second part of Theorem 5,  $\Delta$ is largely determined by the curvature of the function $\log(1+x)$, which  goes to zero when $x\to\infty$. This implies $\Delta$ will increase slowly with the SNR $\rho$.   Indeed, it turns out $\Delta$ has a limit of $\gamma\log e$   as $\rho\to\infty$. For any $\alpha$,  $I(\bm{G}; \bm{X},\bm{Y})/N$ increases with $\rho$ without a limit (refer to (\ref{Corollary2}) for the case $\alpha=1$), then $\Delta$ will be a vanishing fraction of $I(\bm{G}; \bm{X},\bm{Y})/N$ as  $\rho\to\infty$. In that sense, we say that $I(\bm{X};\bm{Y})/N$ is  asymptotically monotonic with respect to $\alpha$ in the high SNR regime. 

We further conjecture that the monotonicity of $I(\bm{X};\bm{Y})/N$ with respect to  $\alpha$ is universal, which means  $I(\bm{X};\bm{Y})/N$ will non-decrease with $\alpha$ for any distribution of $\bm{X}$ and SNR. Since  $I(\bm{X};\bm{Y})=NR_s+I(\bm{G};\bm{Y})-I(\bm{G};\bm{X},\bm{Y})$, if $I(\bm{G};\bm{X}|\bm{Y})=I(\bm{G};\bm{X},\bm{Y})-I(\bm{G};\bm{Y})$ is a non-increasing function of $\alpha$, the conjecture will be true. 
However, since $\bm{Y}$ is not Gaussian in general, a proof through analytic methods seems difficult. More advanced mathematics might be needed to prove the conjecture. Verification through numerical methods is rather practical and expected to come in the near future. 

According to Theorem 4,  $\gamma\log e$ is also an upper bound for  $I(\bm{X};\bm{Y})/N$, the information  rate of an i.i.d. Gaussian channel when the input is also i.i.d. Gaussian. That's about 0.83 bits/symbol in value, which is much lower than the lower bond, about 1.4 bits/symbol at 30dB, achieved by a discrete distribution reported in \cite{G2002Capacity}, because Gaussian distribution is not a capacity approaching one.

\section{Conclusions}
Fading is a common phenomenon in wireless communications.  The total information at the receiver in a fading channel contains a  user message part and a channel information part. The channel information is a non-increasing function of the coherence coefficient $\alpha$ of the channel, meaning more resources will be cost at obtaining the channel information when the fading rate of the channel increases. So we conjecture there exists a monotonic behavior of the information rate of user message with respective to $\alpha$, with an asymptotical monotonicity proved in the case of i.i.d. Gaussian input.  The study of this paper shows that,  channel correlation, in time or frequency domain,  plays a critical role in determining the channel capacity. This philosophy should be extended to multiple antenna systems to examine how the space correlation, hopefully in a similar way as the correlation in the  time and frequency domain,  influences the capacity. With the correction of the misconception of outage capacity, the paper opens up a way to a unified capacity theory for fading channels. 
\section*{Appendix I}

Suppose the joint PDF of $\bm{X},\bm{Y},\bm{G}$ is $ p(\bm{x},\bm{y},\bm{g})$, where 
\begin{equation}\notag\bm{x}=[x_1, x_2,\cdots,x_N]^T,\end{equation} 
\begin{equation}\notag\bm{y}=[y_1, y_2,\cdots,y_N]^T,\end{equation}  
\begin{equation}\notag\bm{g}=\textrm{diag}[g_1, g_2,\cdots,g_N],\end{equation} then
\begin{eqnarray}
I(\bm{G};\bm{X},\bm{Y})  &=&  \int p(\bm{x},\bm{y},\bm{g})\log \frac{p(\bm{x},\bm{y},\bm{g})}{p(\bm{x},\bm{y})p(\bm{g})}d\bm{x}d\bm{y} d\bm{g}\notag\\
   &=&  \int p(\bm{x},\bm{y},\bm{g})\log \frac{p(\bm{y}|\bm{g},\bm{x})}{p(\bm{y}|\bm{x})}d\bm{x}d\bm{y} d\bm{g}. \label{IGXY}
\end{eqnarray}
Notice that the independence of $\bm{X}$ and $\bm{G}$ is used in the above derivation.

Conditioned on $\bm{X}$ and $\bm{G}$, $\bm{Y}$ follows
\begin{eqnarray}\label{y|xg}
p(\bm{y}|\bm{g},\bm{x}) & = & \frac{1}{\pi^N\sigma_Z^{2N}} \exp\left(-\frac{|\bm{y}-\bm{gx}|^2}{\sigma_Z^2}\right).
\end{eqnarray}

Conditioned on $\bm{X}$, $\bm{Y}$ follows
\begin{equation}
\label{y|x}
p(\bm{y}|\bm{x})=\frac{1}{\pi^N\det(\bm{M}_N)}\exp{\left(-\bm{y}^H\bm{M}_N^{-1}\bm{y}\right)},
\end{equation}
where $\bm{M}_N$ is the covariance  matrix of $\bm{Y}$ conditioned on $\bm{X}=\bm{x}$,
\begin{equation}
\label{MYY}
\bm{M}_N= E[ \bm{Y}\bm{Y}^H]=\sigma_Z^2\bm{I}_N+\sigma_{G}^2 \bm{A}_N.
\end{equation}
Putting (\ref{y|xg}) and (\ref{y|x}) into (\ref{IGXY}), we obtain
\begin{equation}
\label{ }
I(\bm{G};\bm{X},\bm{Y})=\int p(\bm{x},\bm{y},\bm{g})\left(\log \frac{ \det(\bm{M}_N)}{\sigma_Z^{2N}}-\frac{|\bm{y}-\bm{gx}|^2}{\sigma_Z^2}+\bm{y}^H\bm{M}_N^{-1}\bm{y}\right)d\bm{x}d\bm{y} d\bm{g}.
\end{equation}

For convenience of narration, we split $I(\bm{G};\bm{X},\bm{Y})$ into three items, with the first one being
\begin{eqnarray}
I_1 &=& \int p(\bm{x},\bm{y},\bm{g})\log \frac{ \det(\bm{M}_N)}{\sigma_Z^{2N}}d\bm{x}d\bm{y} d\bm{g}\notag\\
&=&\int p(\bm{x},\bm{y},\bm{g})\log   \det(\bm{I}_N+\frac{\sigma_G^2}{\sigma_Z^2}\bm{A}_N) d\bm{x}d\bm{y} d\bm{g} \notag\\
&=&  \int p(\bm{x})\log   \det(\bm{I}_N+\frac{\sigma_G^2}{\sigma_Z^2}\bm{A}_N) d\bm{x}.\notag
\end{eqnarray}

The second item
\begin{eqnarray}\label{I2}
I_2 &=&-\int p(\bm{x},\bm{y},\bm{g})\frac{|\bm{y}-\bm{gx}|^2}{\sigma_Z^2}d\bm{x}d\bm{y} d\bm{g}\notag\\
&=&-\int p(\bm{x},\bm{g})p(\bm{y}|\bm{g},\bm{x})\frac{|\bm{y}-\bm{gx}|^2}{\sigma_Z^2}d\bm{x}d\bm{y} d\bm{g}.\label{I2}
\end{eqnarray}
Let $\bm{z}=\bm{y}-\bm{gx}$,  put it and (\ref{y|xg}) into (\ref{I2}), then
  \begin{eqnarray}
  I_2&=& - \int p(\bm{x},\bm{g})p(\bm{z})\frac{|\bm{z}|^2}{\sigma_Z^2} d\bm{z}d\bm{x}d\bm{g}\notag\\&=&- \int p(\bm{z})\frac{|\bm{z}|^2}{\sigma_Z^2} d\bm{z}\notag\\&=&-N,\notag
\end{eqnarray}
where 
\begin{equation}\notag p(\bm{z})=\frac{1}{\pi^N\sigma_Z^{2N} }\exp\left(-\frac{|\bm{z}|^2}{\sigma_Z^2}\right). \end{equation}

The third item   
\begin{eqnarray}
I_3&=&  \int p(\bm{x},\bm{y},\bm{z}) \bm{y}^H\bm{M}_N^{-1}\bm{y}d\bm{x}d\bm{y}d\bm{z}\notag\\
&=&  \int p(\bm{x},\bm{y}) \bm{y}^H\bm{M}_N^{-1}\bm{y}d\bm{x}d\bm{y}\notag\\
&=&  \int p(\bm{x})p(\bm{y}|\bm{x}) \bm{y}^H\bm{M}_N^{-1}\bm{y}d\bm{x}d\bm{y}.\label{I3}
\end{eqnarray}

Perform eigen-decomposition  on $\bm{M}_N=\bm{U\Sigma U}^H$, where $\bm{U}$ is unitary and $\bm{\Sigma}$ is diagonal. The diagonal elements of $\bm{\Sigma}$ are the eigenvalues of $\bm{M}_N$, which are positive real numbers. Let $\bm{y}=\bm{U\sqrt{\Sigma} }\bm{y'}$, then $d\bm{y}=\det(\bm{\Sigma })d\bm{y'}$. Putting them and (\ref{y|x}) into (\ref{I3}), we obtain

\begin{eqnarray}
I_3 &=&  \int p(\bm{x})p(\bm{y'}) |\bm{y'}|^2\frac{\det(\bm{\Sigma})}{\det(\bm{M}_N)} d\bm{y'}d\bm{x}\notag\\
&=& \int p(\bm{y'}) |\bm{y'}|^2d\bm{y'}\notag\\&=& N,\notag
\end{eqnarray}
where \begin{equation}\notag p(\bm{y'})  =  \frac{1}{\pi^N} \exp\left(-|\bm{y'}|^2\right).\end{equation}

Then we have
\begin{eqnarray}
I(\bm{G};\bm{X},\bm{Y}) &=& I_1+I_2+I_3\notag\\&=& \int p(\bm{x})\log   \det(\bm{I}_N+\frac{\sigma_G^2}{\sigma_Z^2}\bm{A}_N) d\bm{x}.
\end{eqnarray}

\section*{Appendix II}

\noindent(1) Since $\alpha=0$, then $\bm{I}_N+\sigma_G^2/\sigma_Z^2\bm{A}_N$ is a diagonal matrix and the proposition is obvious.

\noindent (2) Since $\alpha=1$, then $\bm{A}_N=\bm{xx}^H$, 
\begin{eqnarray}\det(\bm{I}_N+\frac{\sigma_G^2}{\sigma_Z^2}\bm{A}_N)&=&1+\frac{\sigma_G^2}{\sigma_Z^2}\bm{x}^H\bm{x}\notag\\&=&1+\frac{\sigma_G^2}{\sigma_Z^2}|\bm{x}|^2, \end{eqnarray} then we get the conclusion.

\noindent(3) For simplicity, let 
$\beta=\sigma_Z^2/\sigma_G^2,$ $\bm{T}_N=\beta\bm{I}_N+\bm{A}_N,$
then \begin{equation}\bm{I}_N+\frac{\sigma_G^2}{\sigma_Z^2}\bm{A}_N=\bm{T}_N/\beta.\end{equation} 
Let
\begin{eqnarray}
D_N&=&\det(\bm{T}_N)\notag\\
&=&\det\left[                 
  \begin{array}{ccccc}   
   |x_1|^2 +\beta& \alpha x_2 x_1^* & \alpha ^2 x_3 x_1^* & \cdots &\alpha^{N-1} x_N x_1^*\\  
   \alpha x_1 x_2^* &  |x_2|^2+\beta  &  \alpha x_3 x_2^* & \cdots &\alpha^{N-2} x_N x_2^*\\  
    \alpha^2 x_1 x_3^* &  \alpha x_2 x_3^* & |x_3|^2 +\beta  &\ddots &\vdots\\
    \vdots &  \vdots & \ddots  &\ddots &\alpha x_N x_{N-1}^*\\
  \alpha^{N-1} x_1 x_N^* &\alpha^{N-2} x_2 x_N^*&\cdots&\alpha  x_{N-1}x_N^* & |x_N|^2+\beta
  \end{array}
\right].     \label{DNAEDV}            
\end{eqnarray}

If we write down the expansions of $D_1, D_2$ and $D_3$, we have

\begin{equation}\notag
\label{ }
D_1=\beta+ |x_1|^2,
\end{equation}

\begin{equation}\notag
\label{ }
D_2=\beta^2+\beta \sum_{i=1}^2 |x_i|^2+(1-\alpha^2)\prod_{i=1}^2 |x_i|^2,
\end{equation}
\begin{equation}\notag
\label{ }
D_3=\beta^3+\beta^2 \sum_{i=1}^3 |x_i|^2+\beta\sum_{i\neq j}(1-\alpha^{2|i-j|})|x_ix_j|^2+(1-\alpha^2)^2\prod_{i=1}^3 |x_i|^2.
\end{equation}

We actually have proceeded to $D_4$ and $D_5$, whose expansions  are too long to be put in here,  from which we can guess the following general formula

\begin{equation}
\label{DNNRconcise}
D_N=\sum_{i=0}^{N}\beta^{N-i}\sum_{j=1}^{C_N^i}\prod_{k=1}^{i-1}(1-\alpha^{2(s_{j,k+1}-s_{j,k})})\prod_{k=1}^{i}|x_{s_{j,k}}|^2,
\end{equation}
where $(s_{j,1},s_{j,2},\cdots,s_{j,i})$ is the \emph{j}th in the total $C_N^i$ \emph{i}-combinations of the set $(1,2,\cdots,N)$, with the elements ranked in ascending order, i.e. $s_{j,1}<s_{j,2}<\cdots<s_{j,i}$. Here, we adopt the following definition for the product sign, $\prod_{k=i}^{j}a_k=1$ for $i<j$. We now prove (\ref{DNNRconcise}) using induction method.  

It is easy to verify that (\ref{DNNRconcise}) is true for $D_1$ and $D_2$. 

In (\ref{DNAEDV}), if $x_k \neq 0$, it can be extracted  from the $k$th row (row extraction) and multiplied back on the $k$th column (column multiplication), for $k=1,2,\cdots,N$.  Then we have
\begin{equation}\label{Dn}
D_N=\det \left[                 
  \begin{array}{ccccc}   
   |x_1|^2+\beta & \alpha |x_2|^2 & \alpha ^2 |x_3|^2 & \cdots &\alpha^{N-1} |x_N|^2\\  
   \alpha|x_1|^2 &  |x_2|^2+\beta  &  \alpha |x_3|^2 &\cdots &\alpha^{N-2}|x_N|^2\\  
   \alpha^2|x_1|^2 &  \alpha|x_2|^2 &  |x_3|^2+\beta &\cdots &\alpha^{N-3}|x_N|^2\\  
    \vdots &  \vdots & \ddots  &\ddots &\vdots\\
 \alpha^{N-1}|x_1|^2& \alpha^{N-2}|x_2|^2&\alpha^{N-3}|x_3|^2&  \cdots &|x_N|^2+\beta
  \end{array}
\right].                 
\end{equation}

If there is  one  zero element, e.g.  $x_k=0$,  $x_k$ can not be extracted from the $k$th row since $\beta$ can not be divided by zero. In such a case, we  first expand $D_N$ along the $k$th row. Obviously, there is only one non-zero element  located in the $k$th column, which is denoted as $t_{kk}$ and equals $\beta$.  Then $D_N=\beta \det \bm{T}_{N-1}^{kk}$, where $\bm{T}_{N-1}^{kk}$ is the  cofactor of $t_{kk}$.  If there are multiple zero elements, they can be dealt with one by one in  a similar way.  After eliminating the zero elements, we can perform the row extractions and column multiplications on the remaining cofactor to obtain (\ref{Dn}). 

Multiplying the \emph{i}th row of (\ref{Dn}) by $\alpha$ and subtracting it from the (\emph{i}+1)$^{th}$ row, we obtain
\begin{equation}\notag
D_N=\det \left[                 
  \begin{array}{ccccc}   
   |x_1|^2+\beta & \alpha |x_2|^2 & \alpha ^2 |x_3|^2 & \cdots &\alpha^{N-1} |x_N|^2\\  
   -\alpha \beta& (1-\alpha^2) |x_2|^2+\beta  &  \alpha(1-\alpha^2) |x_3|^2 & &\alpha^{N-2}(1-\alpha^2) |x_N|^2\\  
 0  &  -\alpha\beta & (1-\alpha^2) |x_3|^2+\beta    &\ddots &\vdots\\
    \vdots &  \vdots & \ddots  &\ddots &\alpha(1-\alpha^2) |x_N|^2\\
 0&0&\cdots&  -\alpha\beta &(1-\alpha^2) |x_N|^2+\beta
  \end{array}
\right].                
\end{equation}
Expanding $D_N$ along the last row and continuing the expansion recursively, we obtain the following recursive formula
\begin{eqnarray}\label{DNmainRecursive}
D_N &=& [(1-\alpha^2) |x_N|^2+\beta]D_{N-1}+\alpha\beta\stackrel{1}[\alpha(1-\alpha^2) |x_N|^2 D_{N-2}+\notag \\
& & \alpha\beta\stackrel{2}[\alpha^2(1-\alpha^2) |x_N|^2 D_{N-3}+\alpha\beta\stackrel{3}[ \alpha^3(1-\alpha^2) |x_N|^2 D_{N-4}+\alpha\beta\stackrel{4}[\cdots \stackrel{4}]\stackrel{3}]\stackrel{2}]  \stackrel{1}]. 
\end{eqnarray}

Here, the numbers above the  square brackets are used to indicate their levels for readability.

From (\ref{DNmainRecursive}) we can see,  if $x_N=0$, then $D_N=\beta D_{N-1}$.  Therefore, suppose $x_{N-i}=0$ for $i=1,2,\cdots, k-1$, $k<N$, and $x_{N-k}\neq 0$, we have
\begin{equation}
\label{DN1DNK}
D_{N-i} = \beta^{k-i} D_{N-k} ,  i=1,2,\cdots, k-1.
\end{equation} 

Putting (\ref{DN1DNK}) into (\ref{DNmainRecursive}), we obtain
\begin{eqnarray}
D_N &=&\beta^{k}D_{N-k}+(1-\alpha^2) |x_N|^2 \beta^{k-1}D_{N-k}+ \alpha\beta|x_N|^2\stackrel{1}[\alpha(1-\alpha^2) \beta^{k-2} D_{N-k}+\notag \\
& & \alpha\beta\stackrel{2}[\alpha^2(1-\alpha^2)  \beta^{k-3}  D_{N-k}+\alpha\beta\stackrel{3}[\cdots \alpha^{k-1}(1-\alpha^2)  D_{N-k}\stackrel{3}]\stackrel{2}]  \stackrel{1}]+\notag\\
& &\alpha^{k}\beta^{k}|x_N|^2[\alpha^k(1-\alpha^2)  D_{N-k-1}+\alpha\beta[\alpha^{k+1}(1-\alpha^2)  D_{N-k-2}+\alpha\beta[ \cdots ]]\notag\\
&=&\beta^{k}D_{N-k}+(1-\alpha^{2k})|x_N|^2 \beta^{k-1}D_{N-k}+ \label{DNNNK2}\\
& & \alpha^{k}\beta^{k}|x_N|^2[\alpha^k(1-\alpha^2)  D_{N-k-1}+\alpha^k\beta B_{N-k-2}] ,\notag
\end{eqnarray}
where
\begin{eqnarray}
B_{N-k-2} &=& \alpha^2(1-\alpha^2)  D_{N-k-2}+\alpha\beta[ \alpha^3(1-\alpha^2)   D_{N-k-3}+\alpha\beta[\cdots ]]. 
\end{eqnarray}

Refer to (\ref{DNmainRecursive}), we can also expand $D_{N-k} $ as  

\begin{eqnarray}
D_{N-k} &=& [(1-\alpha^2) |x_{N-k}|^2+\beta]D_{N-k-1}+\alpha\beta|x_{N-k}|^2\stackrel{1}[\alpha(1-\alpha^2)  D_{N-k-2}+\notag \\
& & \alpha\beta\stackrel{2}[\alpha^2(1-\alpha^2)  D_{N-k-3}+\alpha\beta\stackrel{3}[ \alpha^3(1-\alpha^2)  D_{N-k-4}+\alpha\beta\stackrel{4}[\cdots \stackrel{4}]\stackrel{3}]\stackrel{2}]  \stackrel{1}]\notag\\
&=& [(1-\alpha^2) |x_{N-k}|^2+\beta]D_{N-k-1}+\beta|x_{N-k}|^2 B_{N-k-2},\label{DNKALPHA}
\end{eqnarray}
then \begin{equation}
\label{BNK2}
B_{N-k-2}=\frac{D_{N-k}-[(1-\alpha^2) |x_{N-k}|^2+\beta]D_{N-k-1}}{\beta|x_{N-k}|^2}.
\end{equation}

Putting (\ref{BNK2}) into (\ref{DNNNK2}), we obtain the following recursive formula
\begin{equation}
\label{DNR}
D_N=\beta^k D_{N-k}+(1-\alpha^{2k})\beta^{k-1}|x_N|^2D_{N-k}+\frac{\alpha^{2k}|x_N|^2}{|x_{N-k}|^2}(\beta^{k} D_{N-k}-\beta^{k+1} D_{N-k-1}).
\end{equation}

Notice (\ref{DNR}) is valid for $N \geqslant k+2$.

Suppose  (\ref{DNNRconcise})  is also true for $D_{N-k}$ and $D_{N-k-1}$, under the condition of $x_{N-i}=0$ for $i=1,2,\cdots, k-1$, we have $D_{N-i} = \beta^{k-i} D_{N-k} ,  i=1,2,\cdots, k-1$. So  (\ref{DNNRconcise}) is true for $D_{N-i},   i=1,2,\cdots, k-1$.

If  $x_{N-i}=0$ for $i=1,2,\cdots, N-2$, which means except $x_1$ and $x_N$, all other elements are zero. In this case,   (\ref{DNR}) is not valid. $D_N$ is directly calculated as
\begin{equation}
\label{ }
D_N=\beta^N+\beta^{N-1} (|x_1|^2+|x_N|^2)+(1-\alpha^{2(N-1)}) |x_1|^2 |x_N|^2.
\end{equation}
So  (\ref{DNNRconcise}) is true  for $D_N$.

If  $x_{N-i}=0$ for $i=1,2,\cdots, k-1$, $k\leqslant N-2$, and $x_{N-k} \neq 0$,  which means there exists at least one non-zero element among $x_2, x_3, \cdots, x_{N-1}$, then we can use  the recursive formula (\ref{DNR}) to verify if  (\ref{DNNRconcise}) is true for $D_N$. 

Let's check if the items in (\ref{DNR}) will match those in (\ref{DNNRconcise}). The items in $\beta^k D_{N-k}$ are all included in (\ref{DNNRconcise}). Among the items in $(1-\alpha^{2k})|x_N|^2D_{N-k}$, those including $|x_{N-k}|^2$ are also correct, but the coefficient $(1-\alpha^{2k})$ is not correct for those without $|x_{N-k}|^2$ and adjustment is necessary. Let's see how this adjustment is performed. 

In the third item of (\ref{DNR}), i.e. $(\beta^{k} D_{N-k}-\beta^{k+1} D_{N-k-1})\alpha^{2k}|x_N|^2/|x_{N-k}|^2$, only the items having $|x_{N-k}|^2$ in $\beta^k D_{N-k}$ are kept by subtracting  $\beta^{k+1} D_{N-k-1}$ from it, and $|x_{N-k}|^2$ is substituted for $|x_{N}|^2$ via multiplying by $|x_N|^2/|x_{N-k}|^2$. Suppose an index set that contains $N-k$ is  $(s_{j,1},s_{j,2},\cdots,s_{j,i}, N-k)$,  the corresponding  coefficient is $\eta (1-\alpha^{2(N-k-s_{j,i})})$,   where $\eta=\prod_{k=1}^{i-1}(1-\alpha^{2(s_{j,k+1}-s_{j,k})})$. Notice $|x_{N-k}|^2$ is substituted for $|x_{N}|^2$, so it is the adjustment in the coefficient for $|x_N|^2\prod_{k=1}^{i}|x_{s_{j,k}}|^2$. The coefficient in $(1-\alpha^{2k})|x_N|^2D_{N-1}$ is $\eta(1-\alpha^{2k})$ for $|x_N|^2\prod_{k=1}^{i}|x_{s_{j,k}}|^2$.   Then the adjusted coefficient is $\eta(1-\alpha^{2k})+\alpha^{2k}\eta (1-\alpha^{2(N-k-s_{j,i})})=\eta(1-\alpha^{2(N-s_{j,i})})$, which is exactly the required  coefficient for  $|x_N|^2\prod_{k=1}^{i}|x_{s_{j,k}}|^2$.

Thus (\ref{DNNRconcise}) is true for $D_N$.

Therefore $D_N$ is a non-increasing function of $\alpha$, and so is $I(\bm{G};\bm{X},\bm{Y})$.

$I(\bm{G};\bm{X},\bm{Y})$ keeps constant with  $\alpha$ if and only if  there is at most one non-zero element in $X_1,X_2,\cdots,X_N$ .

\section*{Appendix III}

Let  \begin{math}
R_l=R_{s}-I(\bm{G};\bm{X},\bm{Y})/N.
\end{math} 

Since $X_1,X_2,\cdots,X_N$ are i.i.d., then  $Y_1,Y_2,\cdots,Y_N$ are independent conditioned on $\bm{G=g}$. Therefore
\begin{eqnarray}
  h(\bm{Y}|\bm{G})&=& \int p(\bm{g}) h(\bm{Y}|\bm{G=g})d\bm{g} \notag\\
 &=& \int p(\bm{g}) \sum_{i=1}^N h(Y_i|G_i=g_i)d\bm{g} \notag\\
 &=& \sum_{i=1}^N \int p(g_i) h(Y_i|G_i=g_i)dg_i=N h(Y_1|G_1) .\notag
 \end{eqnarray}
Hence $R_{s} =[h(\bm{Y}|\bm{G})-h(\bm{Z})]/N=h(Y_1|G_1)-h(Z_1)$  doesn't change with $\alpha$. Since $I(\bm{G};\bm{X},\bm{Y})$  is a non-increasing function of  $\alpha$, then $R_l$ is non-decreasing.

The power of $Y_i$ is $\sigma_G^2\sigma_X^2+\sigma_Z^2$, it's differential entropy can not exceed that of a Gaussian white noise with the same power, then 
\begin{equation}h(\bm{Y})-h(\bm{Z})\leq N\log\left(1+\frac{\sigma_G^2\sigma_X^2}{\sigma_Z^2}\right) =NC_0.\end{equation}
Therefore
\begin{eqnarray}
  I(\bm{G};\bm{Y})&=&h(\bm{Y})-h(\bm{Y}|\bm{G})\notag\\
 &=& h(\bm{Y})-h(\bm{Z})-[h(\bm{Y}|\bm{G})-h(\bm{Z})]\notag\\
 &\leq& N\Delta.
 \end{eqnarray}
Since
\begin{equation}
I(\bm{X};\bm{Y})=h(\bm{Y}|\bm{G})-h(\bm{Z})+I(\bm{G};\bm{Y})-I(\bm{G};\bm{X},\bm{Y})=NR_l+I(\bm{G};\bm{Y}),
\end{equation}
and
\begin{equation}
\label{ }
I(\bm{G};\bm{Y})\geqslant 0, 
\end{equation}
then we have the first part of the theorem. 

If $X_1,X_2,\cdots,X_N$ are i.i.d. and  Gaussian, then
\begin{equation}R_s=\int p(g)\log\left(1+|g|^2\frac{\sigma_X^2}{\sigma_Z^2}\right)dg,\end{equation}
therefore \begin{equation}\Delta= \log\left(1+\frac{\sigma_G^2\sigma_X^2}{\sigma_Z^2}\right)-\int p(g)\log\left(1+|g|^2\frac{\sigma_X^2}{\sigma_Z^2}\right) dg.\end{equation}

According to  \cite{Ericson1970A}, and also \cite{G2002Capacity, 746779}, 
\begin{equation}R_s=(\log e) e^{1/\rho} E_1(1/\rho),\end{equation} 
where
\begin{equation}E_1(x)=\int_x^\infty \frac{e^{-u}}{u}du\end{equation} 
is the exponential integral, which can also be expanded as \cite{Abramowitz1970Handbook}
\begin{equation}E_1(x)=-\gamma-\ln x-\sum_{n=1}^\infty\frac{(-x)^n}{n! n},\end{equation} 
then
\begin{eqnarray}
\Delta&=&\log\left(1+\rho\right)- e^{1/\rho}\log e\left[-\gamma-\ln 1/\rho-\sum_{n=1}^\infty\frac{(-1)^n}{\rho^nn! n}\right]\notag\\
&=&\gamma e^{1/\rho}\log e+\log\frac{1+\rho}{\rho^{\exp(1/\rho)}}+ e^{1/\rho}\log e\sum_{n=1}^\infty\frac{(-1)^n}{\rho^nn! n}. \label{Blimit}
\end{eqnarray} 
The item 
\begin{eqnarray}
\left|\sum_{n=1}^\infty\frac{(-1)^n}{\rho^nn! n}\right|< \sum_{n=1}^\infty\frac{1}{\rho^nn! n}< \sum_{n=1}^\infty\frac{1}{\rho^nn! }=e^{1/\rho}-1.
\end{eqnarray} 
Notice the last two items in (\ref{Blimit}) go to zero when $\rho\to\infty$, then

\begin{equation}\lim_{\rho\to\infty}\Delta=\gamma\log e.  \end{equation} $\Box$

\bigskip\bigskip\bigskip
\bibliographystyle{IEEEtran}
\bibliography{IEEEfull,capacity}

\begin{IEEEbiography}[{\includegraphics[width=1in,height=1.25in,clip,keepaspectratio]{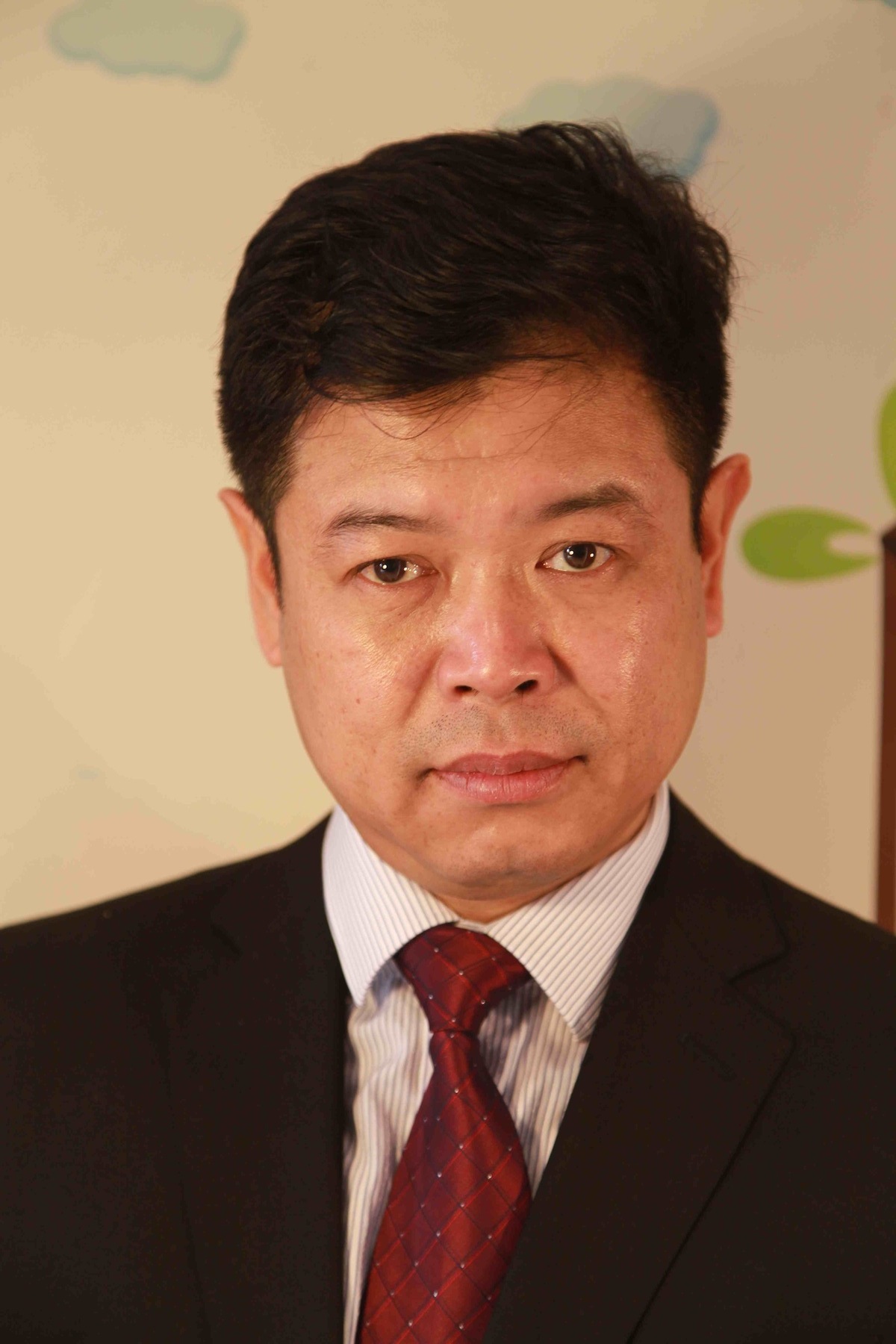}}]{Xuezhi Yang}(SM'12)
was born in 1970, Shandong China. He received the B.E. and Ph.D. degrees in 1993 and 1998 from Tsinghua University Beijing, China. In 1998 he was a Postdoctoral Research Fellow with Peking University, Beijing, China. He was with Huawei Technologies Co., Ltd., China from 2000 to 2012, as a researcher in wireless communications. He is now a freelance researcher on wireless communications.  He has  drafted 50+ patents with 40+ of them granted in China, United States and Europe. He is the inventor of several key technologies of 3G and 4G, including  soft frequency reuse, scalable OFDM, random beamforming and frequency domain multi-user detector.  He is the author of the book \emph{Principles of  Communications}, Publishing House of Electronics Industry, Beijing.  He is an IEEE senior member. His present interests are in information theory,   ICIC, FEC and MIMO. \end{IEEEbiography}

\end{document}